\documentclass{article}

\title{
Restricted trees: simplifying networks with bottlenecks 
}

\author{Stephen J. Willson\\
		Department of Mathematics\\
		Iowa State University\\
		Ames, IA 50011 USA\\
		swillson@iastate.edu}

\usepackage{amsmath}
\usepackage{amsthm}
\usepackage{amssymb}

\usepackage[dvips]{graphicx}

\usepackage{verbatim} 
\newtheorem{lem}{Lemma}[section]
\newtheorem{thm}[lem]{Theorem}
\newtheorem{cor}[lem]{Corollary}

\begin{document}

\maketitle

{\bf Abstract.} 
Suppose $N$ is a phylogenetic network indicating a complicated relationship among individuals and taxa.  Often of interest is a much simpler network, for example, a species tree $T$, that summarizes the most fundamental relationships.  The meaning of a species tree is made more complicated by the recent discovery of the importance of hybridizations and lateral gene transfers.  Hence it is desirable to describe uniform well-defined procedures that yield a tree given a network $N$.  

A useful tool toward this end is a connected surjective digraph (CSD) map $\phi:N \to N'$ where $N'$ is generally a much simpler network than $N$.  A set $W$ of vertices in $N$ is ``restricted'' if there is at most one vertex from which there is an arc into $W$, thus yielding a bottleneck in $N$.  A CSD map $\phi:N \to N'$ is ``restricted'' if the inverse image of each vertex in $N'$ is restricted in $N$.  This paper describes a uniform procedure that, given a network $N$, yields a well-defined tree called the ``restricted tree'' of $N$.  There is a restricted CSD map from $N$ to the restricted tree.  Many relationships in the tree can be proved to appear also in $N$.

\vspace{10 pt}

Key words:    digraph, network, tree, connected, hybrid, phylogeny, homomorphism, restricted, phylogenetic network

\section{Introduction}  

Since Darwin, phylogenetic trees have been utilized to display the evolutionary relationships among taxa.   Extant taxa correspond to the leaves of the trees.  In principle, the trees are directed in the direction of increasing time, and there is a single root indicating the common ancestry of all the taxa in question.  

The underlying reality is often a much more complicated network than a tree.  If every vertex corresponds to an individual and the species are sexually reproducing, then the underlying graph has vast numbers of vertices, each with indegree 2.  This underlying reality is too complicated to reconstruct.  The species phylogenetic tree is a dramatic simplification which summarizes the underlying reality.  

More recently, events such as hybridization and lateral gene transfer have been shown to have increased importance \cite {doo07}, \cite{dag08}.  Such possibilities have called into question the adequacy of  a phylogenetic species tree as a  tool.  Coalescence methods \cite{ros08}, \cite {deg06} have modeled relationships between species trees and gene trees making use of a presumed network of the underlying reality.  Moreover, specific biological networks have been proposed for certain systems \cite{dag08}, \cite{jin07}.  

Once we start to consider networks more general than trees, we must be concerned about the  assumptions that can be made about these networks.  There are astronomically more networks than even the large number of trees with a given leaf set.  Hence it becomes important to narrow the collection in a useful manner.  General frameworks for networks are discussed in \cite{ban92}, \cite{bar04}, \cite{mor04}, \cite{mor09}, and \cite{nak04}.  Typically these frameworks model phylogenies by acyclic rooted directed graphs. 

Particular kinds of networks have been studied in various papers.  Wang {\it et al.} \cite{wan01} and Gusfield {\it et al.} \cite{gel04a} study ``galled trees'' in which all recombination events are associated with node-disjoint recombination cycles.  Van Iersel and others  \cite{ier09} generalized galled trees to ``level-$k$'' networks.  Baroni, Semple, and Steel \cite{bar04} introduced the idea of a ``regular" network, which coincides with its cover digraph.  Cardona {\it et al.} \cite{crv09} discussed ``tree-child" networks, in which every vertex not a leaf has a child that is not a reticulation vertex.  Moret {\it et al.} \cite{mor04} define a reduction $R(N)$ of a network $N$ of use in analyzing displayed trees.  

The possibilities of very complicated networks raise anew the question of the relationship between the hugely complex underlying reality and the phylogenetic trees and networks which simplify and summarize possible relationships. 

Dress {\it et al.}  \cite{dre10} give several abstract constructions of manners in which a very general network can give rise to trees, or, more generally, hierarchies.  For example, they define notions of \emph{tight} clusters and \emph{strict} clusters and show that these produce trees or hierarchies.  Both notions identify a kind of bottleneck in the underlying network  and produce trees.  

In \cite{wil10} the current author described a general approach giving relationships between a complicated underlying network $N$ and a much simpler network $N'$.  For example, $N$ might be the largely unknown directed graph showing the underlying reality while $N'$ might be the species tree.  The basic tool is a \emph{connected surjective digraph} map or, more briefly, a CSD map  $\phi$ from $N$ to $N'$.  The idea is that every vertex $v$ of $N$ is taken to a vertex $\phi(v)$ of $N'$ in such a manner that the following hold:\\
(a) If $(u,v)$ is an arc of $N$, then either $\phi(u) = \phi(v)$ or else $(\phi(u), \phi(v))$ is an arc of $N'$.\\
(b) The map is surjective both on vertices of $N'$ and on arcs of $N'$.\\
(c) For each vertex $v'$ of $N'$, the set of vertices of $N$ mapping to $v'$ forms a connected set.\\
Details are given in section 2.

Many properties of CSD maps are given in \cite{wil10}.  While (a) is very similar to the notion of a homomorphism of digraphs \cite{hah97}, \cite{hel04}, the essential new condition is (c).  Without (c), knowledge of $N'$ gives very little information about $N$; the notion without (c) is too general.  With (c), the notion is much more rigid, and information about  $N'$ implies structure in $N$.  For example, if $N'$ is a binary tree and $\phi: N \to N'$ is a CSD map, then there is a \emph{wired lift} of $N'$ into $N$, showing that as an undirected network $N'$ embeds in $N$.  If, instead, $\phi: N \to N'$ satisfied merely (a) and (b), then when $N'$ is a binary tree, $N$ could still be trivial or a star tree.  Further details are given in section 2 and \cite{wil10}.

The \emph{cluster} of a vertex $v$ in a network $N$ is the set of leaves which can be reached by directed paths starting at $v$.  A network $N$ is \emph{successively cluster-distinct} if whenever $(u,v)$ is an arc, then $u$ and $v$ have distinct clusters.    In \cite{wil10} I gave a construction, given any network $N$, of a successively cluster-distinct network $ClDis(N)$.  I showed that there is a CSD map $\phi: N \to ClDis(N)$, and moreover that $\phi$ had a certain ``universal'' property.  I argued that it was therefore reasonable to restrict one's attention to networks that were successively cluster-distinct.  

In this paper I elaborate further. Given a network $N$, I describe a general method to construct a \emph{restricted} tree denoted $ResTr(N)$.   In some ways the procedure resembles that given in \cite{dre10} of tight clusters in that it detects bottlenecks of a certain sort.  The construction differs, however, in that it always yields a CSD map $\phi: N \to ResTr(N)$; the construction in \cite{dre10} may not have this property.  

The computation of $ResTr(N)$ will typically have more resolution when it is applied to a network $N$ that is already successively cluster-distinct.  

The heart of the construction is the notion of a \emph{restricted} set $B$, given in section 3.  Such a set $B$ is a set of vertices in $N$ such that there is at most one vertex $u$ for which there is any arc $(u,w)$ with $u \notin W$ but $w \in W$.  Such a vertex identifies a bottleneck in the network $N$.  It is shown in Section 3 how to construct the smallest restricted set $R(v)$ containing a given vertex $v$.  These sets are utilized to construct $ResTr(N)$.  

Section 4 focuses on properties of \emph{restricted} CSD maps---those CSD maps for which the inverse images of each point is a restricted set.  It is shown that any such map defined on $N$ factors through $ResTr(N)$, making $ResTr(N)$ ``universal'' for such maps.  Thus $ResTr(N)$ not only permits wired lifts into the network $N$, but any restricted map factors through $ResTr(N)$.   Hence $ResTr(N)$ is an invariantly defined tree with interesting universal properties.

Section 4 also contains an example of the construction of $ResTr(N)$.

\section {Fundamental Concepts} 

A \emph{directed graph} or \emph{digraph} $N=(V,A)$ consists of a finite set $V$ of \emph{vertices} and a finite set $A$ of \emph{arcs}, each consisting of an ordered pair $(u,v)$ where $u \in V$, $v \in V$, $u\neq v$.  Sometimes we write $V(N)$ for $V$.  We interpret $(u,v)$ as an arrow from $u$ to $v$ and say that the arc \emph{starts} at $u$ and \emph{ends} at $v$.  There are no multiple arcs and no loops.   If $(u,v)\in A$, say that $u$ is a \emph{parent} of $v$ and $v$ is a \emph{child} of $u$.  A \emph{directed path} is a sequence $u_0, u_1, \cdots, u_k$ of vertices such that for $i = 1, \cdots, k$, $(u_{i-1}, u_i) \in A$.  The path is \emph{trivial} if $k = 0$.  Write $u \leq  v$ if there is a directed path starting at $u$ and ending at $v$.  Write $u <  v$ if $u\leq v$ and $u \neq v$. The digraph is \emph{acyclic} if there is no nontrivial directed path starting and ending at the same point.  If the digraph is acyclic, it is easy to see that $\leq$  is a partial order on $V$.  

The digraph $(V,A)$ has \emph{root} $r$ if there exists $r \in V$ such that for all $v \in V$, $r\leq v$.  The graph is \emph{rooted} if it has a root.  

The \emph{indegree} of vertex $u$ is the number of $v \in V$ such that $(v,u) \in A$.  The \emph{outdegree} of $u$ is the number of $v \in V$ such that $(u,v) \in A$.  If $(V,A)$ is rooted at $r$ then $r$ is the only vertex of indegree 0.  A \emph{leaf} is a vertex of outdegree 0. A \emph{normal} (or \emph{tree-child}) vertex is a vertex of indegree 1.  A \emph{hybrid} vertex (or \emph{recombination vertex} or \emph{reticulation node}) is a vertex of indegree at least 2.  

Let $X$ denote a finite set.  Typically in phylogeny, $X$ is a collection of species.  An \emph{$X$-network} $N = (V,A,r,X)$ is a 
digraph $(V,A)$ with root $r$ such that \\
(1) there is a one-to-one map 
$\phi: X \to V$ 
such that the image of $\phi$ is the set of all leaves of $(V,A) $, and \\
(2) for every $v \in V$ there is a leaf $u$ and a directed path from $v$ to $u$.\\
Thus the set of leaves of $N$ may be identified with the set $X$; every vertex is ancestral to a leaf. 

In biology most $X$-networks are acyclic.  The set $X$ provides a context for $N$, giving a hypothesized relationship among the members of $X$.  For convenience, we will write $x$ for the leaf $\phi(x)$. 

An \emph{$X$-tree} is an $X$-network such that the underlying digraph is a rooted tree.  

If $N = (V,A,r,X)$ is an $X$-network and $v \in V$, the \emph{cluster} of $v$, denoted $cl(v)$, is $\{x \in X: v\leq x\}$.  We say that $N$ is \emph{successively cluster-distinct} provided that, whenever $(u,v)$ is an arc, then $cl(u) \neq cl(v)$.

Let $N = (V,A,r,X)$ and $N' = (V',A',r',X)$ be $X$-networks.  An \emph{X-isomorphism} $\psi: N \to N'$ is a map $\psi: V \to V'$ such that\\
(1) $\psi: V \to V'$ is one-to-one and onto, \\
(2) $\psi(r) = r'$,\\
(3) for each $x \in X$, $\psi(x) = x$,\\
(4) $(\psi(u), \psi(v))$ is an arc of $N'$ iff $(u,v)$ is an arc of $N$.\\
We say $N$ and $N'$ are \emph{isomorphic} if there is an $X$-isomorphism $\psi: N \to N'$.

A \emph{graph} (or, for emphasis, an \emph{undirected graph}) $(V,E)$ consists of a finite set $V$ of \emph{vertices} and a finite set $E$ of \emph{edges}, each consisting of a subset $\{v_1,v_2\}$ where $v_1$ and $v_2$ are  two distinct members of $V$.  Thus an edge has no direction, while an arc has a direction.    If $G=(V,E)$ is a graph and $W$ is a subset of $V$, the \emph{induced subgraph} $G[W]$ is the graph $(W,E[W])$ where the edge set $E[W]$ is the collection of all $\{v_1,v_2\}$ in $E$ such that $v_1 \in W$ and $v_2 \in W$.  Thus $G[W]$ contains all edges both of whose endpoints are in $W$. 

A graph $G=(V,E)$ is \emph{connected} if, given any two distinct $v$ and $w$ in $V$ there exists a sequence
$v = v_0, v_1, v_2, \cdots, v_k = w$ of vertices  such that for $i = 0, \cdots, k-1$, $\{v_i, v_{i+1}\}\in E$.  
A subset $W$ of $V$ is \emph{connected} if the induced subgraph $G[W]$ is connected.

Given a digraph $G=(V,A)$ define $Und(G) = (V,E)$ where $E = \{\{u,v\}:$ there is an arc $(u,v) \in A\}$.  Then $Und(G)$ is an undirected graph with the same vertex set as $G$ and with edges obtained by ignoring the directions of arcs.  A subset $W$ of $V$ is \emph{connected} if $Und(G)[W]$ is connected.   Thus a connected subset of $G$ is defined ignoring the directions of arcs.

Let $N=(V,A,r,X)$ and $N'=(V',A',r',X)$ be  $X$-networks whose leaf sets are identified with the same set $X$.  An \emph{X-digraph map}   $f: N\to N'$
is a map $f: V \to V'$ such that\\
(a) $f(r) = r'$,\\
(b) for all $x \in X$, $f(x) = x$, and \\
(c) if $(u,v)$ is an arc of $N$, then either $f(u) = f(v)$ or else $(f(u), f(v))$ is an arc of $N'$.

Call $f$ \emph{connected} if for each $v' \in V'$, $f^{-1}(v')$ is a connected subset of $N$, {\it i.e.}, if the induced subgraph $Ind(N)[f^{-1}(v')]$ is  connected.
Call $f$ \emph{surjective} if for each $v' \in V$, $f^{-1}(v')$ is nonempty and for each arc $(a,b)$ of $N'$ there exist vertices $u$ and $v$ of $N$ such that $(u,v)$ is an arc of $N$, $f(u) = a$, and $f(v) = b$.  
The \emph{kernel} of $f$ is the partition $\{\{f^{-1}(v')\} :v' \in V'\}$ of $V$.  

We are interested primarily in $X$-digraph maps that are both connected and surjective.  They will be called \emph{connected surjective digraph maps}  or \emph{CSD maps}.  Many of their properties are analogous to properties of homomorphisms \cite{hel04} but properties involving the leaf set $X$ and connectivity require special attention. 

The following basic results are in the paper \cite{wil10}.

Let $N = (V,A,r,X)$ be an  $X$-network.  If $\sim$ is an equivalence relation on $V$, denote by $[v]$ the equivalence class of the vertex $v$.   An equivalence relation $\sim$ on $V$ is called \emph{leaf-preserving} provided that  for every $x \in X$ whenever $u \in [x]$ and $(u,v)$ is an arc, then $v \in [x]$. 

Let $N=(V,A,r,X)$ be an $X$-network.  Suppose $\sim$ is an equivalence relation on $V$.   Let $\mathcal{P} = \{[v]: v \in V\}$ be the partition of $V$ into equivalence classes.  Define the \emph{quotient digraph} $N'$   by $N' = (V',A',r',X)$ where \\
(i) $V'$ is the set of equivalence classes $[v]$.\\
(ii) $r' = [r]$.\\
(iii) The member $x\in X$ corresponds to $[x]$; {\it i.e.}, the identification is given by $\phi': X \to V'$ by $\phi'(x) = [\phi(x)]$.\\
(iv) Let $[u]$ and $[v]$ be two equivalence classes.  There is an arc $([u], [v]) \in A'$ iff
$[u] \neq [v]$  and there exists $u' \in [u]$ and $v' \in [v]$ such that $(u',v')\in A'$.\\
Alternative notations for $N'$ will be  $N/\sim$ or $N/\mathcal{P}$.  

\begin{thm}
Let $N = (V,A,r,X)$ be an  $X$-network.  Suppose $\sim$ is a leaf-preserving equivalence relation on $V$ .  Let $N' = N/\sim  \:= (V',A',r',X)$ be the quotient digraph. Then \\
(1) $N'$ is an $X$-network.\\
(2) The natural map $\phi: N \to N'$ 
given by $\phi(u) = [u]$
is a surjective $X$-digraph map with kernel the set of equivalence classes under $\sim$.\\
(3)  If each equivalence class $[u]$ is connected in $N$, then $\phi$ is connected.
\end{thm}

\begin{thm}
Let $N = (V,A,r,X)$ and $N' = (V',A',r',X)$ be $X$-networks. 
Suppose $f: N \to N'$ is a surjective $X$-digraph map.  Define the relation $\sim$ on $V$  by $u\sim v$ iff $f(u) = f(v)$.   Then $\sim$ is a leaf-preserving equivalence relation and the equivalence classes are $[u] = f^{-1}(f(u))$.  Moreover the quotient digraph $N/\sim$ is isomorphic with $N'$ via the map $\phi: N/\sim \: \to N'$ given by
$\phi ([u]) = f(u)$.
\end{thm}

\begin{thm}
Let $N$ and $N'$ be $X$-networks.  Let $f: N \to N'$ and $g: N' \to N''$ be $X$-digraph maps.\\
(a) The composition $g \circ f : N \to N''$ is an $X$-digraph map.\\
(b)  If $f$ and $g$ are surjective, then $g \circ f$ is surjective.\\
(c) If $f$ and $g$ are connected and surjective, then $g \circ f$ is connected and surjective.
\end{thm}

Suppose $N=(V,A,r,X)$ is an $X$-network.  
A partition $\mathcal{Q}$ of $V$ is \emph{subordinate} to a partition $\mathcal{P}$ of $V$ provided, for each $A \in \mathcal{Q}$, there exists $B \in \mathcal{P}$ such that $A \subseteq B$.  

\begin{thm}
Let $N=(V,A,r,X)$ and $N'=(V',A',r',X)$ be $X$-networks.  
Let $f: N \to N'$ be a surjective $X$-digraph map with kernel $\mathcal{P} = \{f^{-1}(v): v \in V'\}$.   Suppose $\mathcal{Q}$ is a partition of $V$ that is subordinate to $\mathcal{P}$.  \\
(1) There exist surjective $X$-digraph maps $g: N \to N/\mathcal{Q}$ and $h: N/\mathcal{Q} \to N'$ such that 
$f = h \circ g$.\\
(2) If in addition $f$ is connected and each member of $\mathcal{Q}$ is connected, then both $h$ and $g$ are connected.
\end{thm}

Let $N = (V,A,r,X)$ and $N' = (V',A',r',X)$ be $X$-networks.  Suppose $f: N \to N'$ is a surjective digraph map.  A \emph{wired lift} of $N'$ is a subgraph $M =(W,E)$ of $Und(N)$ such that the following hold:\\
(1) For each arc $(u',v')$ of $N'$ there is exactly one arc $(u,v)$ of $N$ such $f(u) = u'$, $f(v) = v'$, and $\{u,v\}$ is an edge of $M$.   The set of all edges $\{u,v\}$ so obtained will be denoted $E_1$ and the set of all vertices which occur in any of the arcs $(u,v)\in E_1$ will be denoted $V_1'$.  Let $V_1 = V_1' \cup X$.\\
(2) Every edge $\{a,b\} \in E$ either lies in $E_1$ or else satisfies $f(a) = f(b)$.\\
(3)  For each vertex $u'$ of $N'$,  let $V(v') = \{w \in V_1: f(w) = v'\}$.  The induced subgraph $M[  f^{-1}(u') \cap W]$ is a tree with leafset $V(v')$. 
 
We call $E_1$ the set of \emph{nondegenerate} edges of $M$, since the image under $f$ of each such edge is an edge of $N'$, not just a single vertex.  Note that $W \subseteq V$ and $E\subseteq E(Und(N))$.  

Intuitively, $M$ is a subgraph of $Und(N)$ that is a resolution of $Und(N')$ in that for each vertex $v'$ of $N'$, $[f^{-1}(v')] \cap W$ consists of the vertices of a tree, all of whose vertices map to $v'$, not necessarily a single point.  The name ``lift'' suggests that $N'$ is being lifted into the domain of $f$.  

The following theorem gives sufficient conditions for a wired lift to exist given any choice of $E_1$.  The essential property is that $f$ be connected.  

\begin{thm}
Let $N = (V,A,r,X)$ and $N' = (V',A',r',X)$ be $X$-networks.  Suppose $f: N \to N'$ is a CSD map.  For each arc $(u',v')$ of $N'$ choose an arc $(u,v)$ of $N$ such that $\phi(u) = u'$, $\phi(v) = v'$. Let $E_1$ denote the set of edges $\{u,v\}$ of $Und(N)$ so obtained.  Then $f$ has a wired lift $M$ for which $E_1$ is the set of nondegenerate edges.  Each such wired lift $M$ is a resolution of $Und(N')$.  
\end{thm}

\section{Restricted sets}

Let $N = (V,A,r,X)$ be a rooted acyclic $X$-network.  We seek natural methods to assign standard networks of various sorts  to $N$.  For example, even if $N$ has many hybridization events, we might be able to assign some standard tree that might correspond to some consensus species tree.   

This section proposes one such construction, which will be denoted $ResTr(N)$.  
An example is given in Section 4.  
$ResTr(N)$ will have the form $N/\sim$ for a certain equivalence relation $\sim$  on the vertices of $N$.  Because of the construction, there will be a CSD map $f:N \to ResTr(N)$.  Consequently, by Theorem 2.5, $ResTr(N)$ will have a wired lift into $N$.  

In this section we shall assume that $N = (V,A,r,X)$ is a rooted acyclic network with leaf set $X$.  We shall sometimes assume that every leaf is tree-child (with indegree 1). 

The construction involves identifying subsets of $V$ here called ``restricted subsets."  

A set $B$ of vertices is called \emph{closed} if, whenever $b_1$ and $b_2$ are in $B$ and $b_1 < b_2$, then every vertex $v$ such that $b_1 < v < b_2$ also lies in $B$.   

A nonempty set $B$ of vertices not containing $r$  \emph{has restricted entry} or is \emph{restricted} if there exists a unique vertex $w'$ such that \\
(1) $w' \notin B$, \\
(2) for some $b \in B$ there is an arc $(w',b)$,  \\
(3) whenever $(w,b)$ is an arc, $w \notin B$, $b \in B$, then $w = w'.$  \\
We call this unique vertex $w'$ the \emph{anchor} of $B$ and write $Anc(B) = w'$.  A set $B$ of vertices containing $r$ \emph{has restricted entry} or is \emph{restricted} if  there is no arc $(w,b)$ with $b \in B$ and $w \notin B$.  

\begin{lem}
A restricted set $B$ is closed.  
\end{lem}

\begin{proof}
Suppose first that $B$ does not contain $r$.  Suppose $b_1 <  v <  b_2$ with $b_1 \in B$, $b_2 \in B$, $v \notin B$.   We may assume $(v, b_2)$ is an arc, whence $v = Anc(B)$.   But since $r$ is the root, there is a directed path $P$ from $r$ to $b_1$; since $r \notin B$ and $b_1 \in B$,  $Anc(B)$ lies on $P$.  It follows $Anc(B) <  b_1 <  Anc(B)$, so that $N$ has a directed cycle, contradicting that $N$ is acyclic.  

To see that $B$ is closed if $B$ contains $r$, suppose $b_1 <  v <  b_2$ with $b_1 \in B$, $b_2 \in B$, $v \notin B$.  We may assume $(v, b_2)$ is an arc, contradicting that $B$ is restricted.  
\end{proof}

\begin{lem}
Let $N=(V,A,r,X)$ be an acyclic $X$-network.  Suppose $B$ is restricted and $r \notin B$.  For every $b \in B$ there is a directed path from $Anc(B)$ to $b$ such that all vertices on the path except $Anc(B)$ itself lie in $B$.
\end{lem}

\begin{proof}
Choose a path from $r$ to $b$, say $r = u_0$, $u_1$, $\cdots$, $u_k = b$.  Since $r \notin B$ and $u_k \in B$, there exists $i$ such that $u_i \notin B$, $u_{i+1} \in B$.   Since $B$ has an anchor, it follows $u_i = Anc(B)$.  Since $u_{i+1} \in B$ and $u_k \in B$, every vertex on the path from $u_{i+1}$ to $u_k$ lies in $B$ because $B$ is closed by Lemma 3.1.   \end{proof}

\begin{thm}
Let $N=(V,A,r,X)$ be an acyclic $X$-network.   Suppose $B$ and $C$ are restricted subsets and $B \cap C$ is nonempty.  Then $B \cup C$ is restricted.
\end{thm}

\begin{proof}
 Assume $w \in B \cap C$.  We prove the result via three cases.

Case 1.  Suppose $r$ is in neither $B$ nor $C$.  Then both $B$ and $C$ have anchors.  I claim first that either $Anc(B) \in C$ or $Anc(C) \in B$ or $Anc(B) = Anc(C)$.  To see this, suppose $Anc(B) \notin C$.  Since $w \in B$ by Lemma 3.2 there is a directed path $P$ from $Anc(B)$ to $w$  such that all vertices after the first lie in $B$.   Since $Anc(B) \notin C$, there is a vertex $v$ on the path $P$ which is not in $C$ but whose child on the path lies in $C$.  Hence $v = Anc(C)$.  It follows that either $Anc(C) = Anc(B)$ or else $Anc(C) \in B$.   This proves the claim.

Now there are three subcases:\\
Subcase (1a).  Suppose $Anc(B) \in C$.  

To show that $B \cup C$ is restricted, since $r \notin B \cup C$, it suffices to show that $Anc(C)$ is an anchor for $B \cup C$.  To see this, suppose $(u,d)$ is an arc with $d \in B \cup C$ and $u \notin B \cup C$.  If $d \in C$, then $u = Anc(C)$.  If $d \in B$, then $u = Anc(B)$, but this implies $u \in C$, contradicting that $u \notin B \cup C$; so this latter case cannot occur.

Subcase (1b)  Suppose $Anc(C) \in B$.   Then $Anc(B)$ is an anchor for $B \cup C$ and $B \cup C$ is restricted by arguments like those in subcase (1a). 

Subcase (1c) Suppose $Anc(B) = Anc(C)$.  I claim $Anc(B)$ is an anchor for $B \cup C$.  To see this, suppose $(u,d)$ is an arc with $d \in B \cup C$ and $u \notin B \cup C$.  If $d \in B$ then $u = Anc(B)$.  If $d \in C$ then $u = Anc(C) = Anc(B)$.

Hence the result is true in Case 1.  

Case 2.  Suppose $B$ has an anchor but $r \in C$.  I claim $B \cup C$ is restricted.   Since $r \in B \cup C$, we suppose $(u,d)$ is an arc with $u \notin B \cup C$ but $d \in B \cup C$, and we derive a contradiction.  Since $C$ is restricted and contains $r$, it follows $d \notin C$.  Hence $d \in B$ and $u = Anc(B)$.  

Since $B$ has an anchor and $w \in B$, by Lemma 3.2 there is a path from $Anc(B)$ to $w$ such that all vertices after the first lie in $B$.  Since $r$ is the root, we obtain a path from $r$ to $Anc(B)$ and then to $w$.  Since $w \in C$ and $C$ is closed by Lemma 3.1, it follows $Anc(B) \in C$.  This contradicts that $Anc(B) = u \notin B \cup C$.  Hence the result is true in Case 2.  

Case 3.  Suppose $r \in B$ and $r \in C$.  I claim $B \cup C$ is restricted.  Since $r \in B \cup C$ we suppose $(u,d)$ is an arc with $u \notin B \cup C$ but $d \in B \cup C$, and we derive a contradiction. Note that we cannot have $d \in B$ since $B$ is restricted, and we cannot have $d \in C$ since $C$ is restricted.  Hence the situation is not possible.
\end{proof}

Another way to combine restricted sets into a new restricted set is given in the next result:

\begin{lem}
Suppose $B$ and $C$ are restricted sets and there is an arc $(b,c)$ with $b \in B$ and $c \in C$.  Then $B \cup C$ is restricted.  
\end{lem}

\begin{proof}
If $B$ and $C$ intersect, then the result follows from Theorem 3.3.  So we may assume that $B$ and $C$ are disjoint.   Since $b \notin C$ and $C$ is restricted, it follows that $b = Anc(C)$.  Now suppose that $(u,v)$ is an arc with $v \in B \cup C$ and $u \notin B \cup C$.  If $v \in C$, then $u = Anc(C)$, so $u \in B$, a contradiction.  Hence $v \in B$, so $r \notin B$ and $u = Anc(B)$.  Since $u$ is uniquely determined, $B \cup C$ is restricted.   
\end{proof}

Now, whenever $v \in V$, we construct an interesting restricted set denoted  $R(v)$.  It will turn out that $R(v)$ is the smallest restricted set that contains $v$.  

The basic construction is the following:  
\bigskip
\hrule height1pt
\medskip
\noindent\textbf{Algorithm} Smallest restricted set\\
\textbf{Input}.  An acyclic $X$-network N = $(V,A,r,X)$ and $v \in V$.\\
\textbf{Output}.  A subset $R(v)$ of $V$.\\
\textbf{Procedure}: Define a sequence of sets $R_i$ of vertices as follows:\\
(1) Let $R_0 = \{v\}$.\\
(2) Recursively, given $R_i$ perform the following:  
Suppose there exist $u \notin R_i$ and $w \in R_i$ with arc $(u,w)$.  Let $R_{i+1} := R_i \cup \{u\}$ if either of the following holds:\\
(a) there exists $w' \in R_i$ such that $u \nleq  w'$;\\
(b) there exist $u' \in V - R_i$, $v' \in R_i$, $u' \neq u$, and arc $(u',v')$ such that  $u \nleq Êu'$. \\
(3) Iterate the procedure until for some $m$, $R_m$ has been constructed and there are no further changes possible according to (2).  Define $R(v) = R_m$.  
\medskip
\hrule height1pt
\bigskip

An example of the algorithm is given in section 4.

In step (2), if there are two vertices $u$ and $u'$ not in $R_i$ , $u\neq u'$, and arcs $(u,v')$, $(u',v'')$ with $v'$ and $v''$ in $R_i$, then we cannot have both $u \leq  u'$ and $u' \leq  u$ since that would force $u = u'$.  Hence  at least one of $u$ and $u'$ will be adjoined to $R_i$.  It is possible that both $u$ and $u'$ will be adjoined to $R_i$ in separate steps.  

It is easy to see that $R(v)$ is well-defined.   This assertion means that when the algorithm terminates,  the result $R(v)$ is independent of the order in which the operations were carried out as long as they were legitimate when performed.  

To see this, suppose at a certain time we have $u_1$, $u_2$ not in $R_i$, $v_1 \in R_i$, $v_2 \in R_i$, $u_1 \neq u_2$, and arcs $(u_1,v_1)$, $(u_2,v_2)$.  If there exists $w \in R_i$ and  $u_1 \nleq  w$, we could adjoin $u_1$.   Alternatively if we are able to adjoin $u_2$ first and then consider $u_1$, it is still true that $w \in R_i$ and  $u_1 \nleq  w$, so we can still adjoin $u_1$.  Another possible scenario is that $u_1 \nleq  u_2$ and  $u_2 \nleq  u_1$, so either could be adjoined first.  Then we may adjoin $u_1$ and at a later stage $u_2$ still meets the criterion for adjoining $u_2$ since now $w' = u_1$ applies for (2a).  

Note that if $u$ has indegree 1, then $R(u) = \{u\}$ since no operation of type (2) can be carried out.  

\begin{thm}
Let $N=(V,A,r,X)$ be an acyclic $X$-network.  For each $v \in V$, $R(v)$ is restricted.
\end{thm}

\begin{proof}
Suppose first that $r \in R(v)$.  We must show that there is no vertex $w$, $w \notin R(v)$, such that there is an arc $(w,b)$ with $b \in R(v)$.  Otherwise, if such $w$ exists, then there is a path from $r$ to $w$ then to $b$ with $w \notin R(v)$. Note that $R(v) = R_m$ for some $m$.  Moreover $w \nleq  r$ since this can happen only when $w = r$ and $w \notin R_m$ but $r \in R_m$.   Hence step (2a) could be used to define $R_{m+1} := R_m \cup \{w\}$, contrary to the assumption that no more operations of type (2) can be performed. 

Now suppose that $r \notin R(v)$.  We show that $R(v)$ has an anchor.  Since $r \notin R(v)$ there is a path from $r$ to some member $b \in R(v)$ and a vertex $w$ on the path which is not in $R(v)$ but such that the next vertex on the path lies in $R(v)$.  This proves there exists $w \notin R(v)$ and an arc $(w, b)$ with $b \in R(v)$.  To have an anchor, this vertex $w$ must be unique, in which case $R(v)$ is restricted.  Suppose there were two vertices $w_1$ and $w_2$ with arcs $(w_1,b_1)$ and $(w_2, b_2)$, $w_1\neq w_2$, $w_1 \notin R(v)$, $w_2 \notin R(v)$, $b_1 \in R(v)$, $b_2 \in R(v)$.  Note that $R(v) = R_m$ for some $m$.   If $w_1 \nleq  w_2$ then step (2b) could be used to enlarge $R_m$ by adjoining $w_1$, and similarly if $w_2  \nleq  w_1$ then $R_m$ could be enlarged by adjoining $w_2$.  Hence $w_1 \leq  w_2$ and $w_2 \leq Êw_1$, implying $w_1 = w_2$.  This proves that the vertex $w$ is unique, so $R(v)$ is restricted.  
\end{proof}

The sets $R(v)$ have other nice properties.  The next result shows that $R(v)$ is the smallest restricted set that contains $v$.  

\begin{thm} 
Let $N=(V,A,r,X)$ be an acyclic $X$-network.  Suppose $B$ is a restricted set and $w \in B$.  Then $R(w) \subseteq B$.
\end{thm}

\begin{proof}
Let the sequence $R_i$ be used to compute $R(w)$.  Initially $R_0 = \{w\} \subseteq B$.   The proof will be by induction.  We will assume $R_i \subseteq B$ but the algorithm does not terminate with $R_i$.  We will prove $R_{i+1} \subseteq B$.  It is immediate that $R_0 \subseteq B$.  

Note that $R_{i+1}$ arises from $R_i$.  Hence there exist $u \notin R_i$, $v \in R_i$, and arc $(u,v)$  such that $u$ is adjoined to $R_i$ in one of two ways.  We must show that $u\in B$. 

Suppose (2a) applies.  Hence there exists $w' \in R_i$ such that $u  \nleq  w'$; we show $u \in B$.  If not, then since $v \in B$ and $B$ is restricted, it follows $u = Anc(B)$.  Since $w' \in R_i$, we have $w' \in B$  since $R_i\subseteq B$, whence by Lemma 3.2, $u\leq w'$, a contradiction.  Hence $u \in B$ so $R_{i+1} \subseteq B$.

Suppose instead (2b) applies.  Hence there exist  $u' \notin R_i$,  $u\neq u'$, $v' \in R_i$, and arc  $(u',v')$, such that $u  \nleq  u'$.  We show $u \in B$.  If not, then $u = Anc(B)$ since $v \in B$.  We cannot have $u' \notin B$, since then $u' = Anc(B) = u$.  Hence $u' \in B$, whence by Lemma 3.2, $u = Anc(B) \leq  u'$, a contradiction.  This proves $u \in B$ so $R_{i+1}\subseteq B$.
\end{proof}

\begin{cor}
If $u \in R(v)$, then $R(u) \subseteq R(v)$.
\end{cor}

\begin{proof}
By Theorem 3.4, $R(v)$ is restricted.  The result follows now from Theorem 3.6.
\end{proof}

In fact, given any subset $B$ of $V$, the algorithm computes the smallest restricted set that contains $B$ provided that we use $R_0 = B$.

In general it need not be the case that a restricted set $B$ is connected.  For example, suppose that $N$ is an $X$-tree and the leaves $x$ and $y$ form a cherry, so there are a vertex $u$ and arcs $(u,x)$, $(u,y)$ with no other arcs into $x$ or $y$.  Then $\{x,y\}$ is restricted with anchor $u$ but is not connected.  Consequently, the following result that each set $R(v)$ is connected is of interest.

\begin{lem}
For $v \in V$, if $w \in R(v)$, then there is a directed path in $R(v)$ from $w$ to $v$.  Moreover, $R(v)$ is connected.
\end{lem}
\begin{proof}
We show the properties for each $R_i$ used to define $R(v)$.  Initially  $R_0 = \{v\}$ and the properties are immediate.  Each operation (2a) or (2b)  applied to $R_i$ results in a connected set $R_{i+1}$ and adds a vertex with a path to $v$ inside $R_{i+1}$.  
\end{proof}

Let $\mathcal{Q}  = \{ R(v): v \in V\}$.   Note that  $\mathcal{Q}$ does not need to be a partition of $V$, but each $v \in V$ lies in at least one member of $\mathcal{Q}$.  

We now make some modifications of $\mathcal{Q}$ to create a partition $\mathcal{P}$ of $V$.  Roughly we merge together members of $\mathcal{Q}$ that have nonempty intersection until no more such merges can be performed.  

More precisely, if $R(v)$ and $R(w)$ are in $\mathcal{Q}$, define $R(v) \sim  R(w)$ if $R(v) \cap R(w) \neq \emptyset$.  Define $R(v) \approx R(w)$ iff there exist $v=v_0$, $v_1$, $\cdots$, $v_k=w$ such that  for $i = 0$, $\cdots$, $k$, $R(v_i) \in \mathcal{Q}$ and for $i = 0$, $\cdots$, $k-1$, $R(v_i) \sim  R(v_{i+1})$.  Then $\approx$ is an equivalence relation.   Define
$R'(v) = \cup \{ R(w): R(v) \approx R(w)\}$,  so $R'(v)$ is the union of sets equivalent to $R(v)$. 
Let $\mathcal{P} = \{R'(v)\}$ be the set of distinct sets $R'(v)$.  It is clear that $\mathcal{P}$ is a partition of $V$. For $v \in V$, $R'(v)$ is the member of $\mathcal{P}$ containing $v$. 

\begin{lem}
Each set $R'(v)$ is a  restricted subset of $V$ and is connected.  
\end{lem}
\begin{proof} The fact that $R'(v)$ is restricted follows from Theorems 3.3 and 3.5 by an obvious induction.  That $R'(v)$ is connected follow from a similar induction, also using Lemma 3.8.  \end{proof} 

Let $ResTr(N) = N / \mathcal{P}$ be the quotient $X$-network.   The map $\phi :N \to ResTr(N)$  given by $\phi(v) = R'(v)$ will be called the \emph{natural projection map}.  

\begin{thm}
Suppose $N = (V,A,r,X)$ is a rooted acyclic network with leaf set $X$ such that every leaf has indegree 1.  Then $ResTr(N)$ is an $X$-network.  The natural projection map $\phi: N \to ResTr(N)$ is a CSD map. 
\end{thm}

\begin{proof}
For $x \in X$, since $x$ has indegree 1, it follows $R(x) = \{x\}$.  If $v$ is not a leaf, then each $w \in R(v)$ satisfies $w \leq  v$ by Lemma 3.8; it follows that a leaf $x$ cannot lie in $R(v)$ when $v$ is not a leaf.    Hence $R'(x) = \{x\}$.  By Theorem 2.1, it follows that $ResTr(N)$ is a rooted digraph with leaf set $X$.  By  Theorem 2.1 and Lemma 3.9, the natural projection map $\phi: N \to N'$ is a CSD map.  
\end{proof}

It will turn out (Theorem 4.1) that $ResTr(N)$ is an $X$-tree, and we will call it the  \emph{(standard) restricted tree} of $N$. The corresponding kernel $\mathcal{P}$ will be called the \emph{restricted tree kernel}.

\section{Restricted maps} 

A CSD map $f: N \to N'$ with kernel $\mathcal{Q}$ is \emph{restricted} if each member of $\mathcal{Q}$ is restricted.  Equivalently, $f$ is restricted if for each vertex $v'$ of $N'$, $f^{-1}(v')$ is a restricted set.  

The natural projection map $\phi: N \to ResTr(N)$ is a restricted map since each member of the kernel $\mathcal{P}$ is a restricted set.  

Suppose a network $N$ is successively cluster-distinct.  Then a restricted set $B$ is a natural generalization of a taxon unit in a tree.  Each restricted set $B$ corresponds to a connected collection of taxa all deriving from the single taxon $Anc(B)$.  If $N$ is a tree, then each vertex is already restricted; the image of a restricted map thus generalizes the notion of a tree.  Note that \cite{dre10} argues that the extant human population forms a tight cluster.  If $N$ is successively cluster-distinct, the same argument would suggest that it forms a restricted set.  

A restricted CSD map $f: N \to N'$ is \emph{universal (for restricted maps)} provided that given any restricted map $g: N \to N''$  there is a unique restricted CSD map $h: N' \to N''$ such that $g = h \circ f$.

We shall see below that the natural projection map $\phi: N \to ResTr(N)$ is universal for restricted maps.  

The first result is that the image of a restricted map is always a tree.

\begin{thm}
Let $N=(V,A,r,X)$ be an acyclic $X$-network and let $T = (V',A',r',X)$ be an $X$-network.  Assume $f: N \to T$ is a restricted CSD map.  Then $T$ is a tree.  
\end{thm}

\begin{proof} 
We show that $T$ has no hybrid vertices.  Suppose otherwise, so we may assume $V'$ contains distinct vertices $u_1'$, $u_2'$, and $u_3'$ while $A'$ contains arcs $(u_1',u_3')$, $(u_2', u_3')$.  Let $B_i = f^{-1}(u_i')$.  Since $f$ is restricted, each $B_i$ is a restricted set.  Since $f$ is a CSD map, there exist $u_1 \in B_1$, $w_1 \in B_3$, $u_2 \in B_2$, and  $w_2 \in B_3$ such that $(u_1,w_1)$ and $(u_2,w_2)$ are arcs of $N$.  Note $u_1 \notin B_3$ and $u_2 \notin B_3$.  Since $B_3$ is restricted, it follows $u_1 = Anc(B_3)$ and $u_2 = Anc(B_3)$.  Hence $u_1 = u_2$ so $u_1' = f(u_1) = f(u_2) = u_2'$, a contradiction.  
\end{proof}

\begin{cor}
Let $N=(V,A,r,X)$ be an acyclic $X$-network.  Then $ResTr(N)$ is an $X$-tree.    
\end{cor}

\begin{proof}
The natural projection map $\phi: N \to ResTr(N)$ is a restricted CSD map.  
\end{proof}

The next result shows that many relationships among leaves observed in $ResTr(N)$ are also present in $N$.  

\begin{cor}
Let $N=(V,A,r,X)$ be an acyclic $X$-network.  There is a wired lift of $ResTr(N)$ into $N$.   
\end{cor}

\begin{proof}
This follows from Theorem 2.5.
\end{proof}

Restricted maps have interesting functorial properties, as seen in the next results.

\begin{lem}
Suppose $N=(V,A,r,X)$ and $N' = (V',A',r',X)$ are $X$-networks.  Suppose $f:N \to N'$ is restricted.   If $B \subseteq V'$ is restricted and connected, then $f^{-1}(B)$ is restricted.
\end{lem}

\begin{proof}
Let $B = \{w_1', w_2', \cdots, w_m'\} \subseteq V'$.  Since $B$ is  connected, for some $p$ there exist $p$ arcs $(w_{i_1}',w_{j_1}')$, $\cdots$, $(w_{i_p}',w_{j_p}')$  such the arcs connect the members of $B$.  Now each set $f^{-1}(w_i')$ is restricted.  Since $f$ is a CSD map for each $k$ there is an arc $(w_{i_k},w_{j_k})$ with $w_{i_k} \in f^{-1}(w_{i_k}')$ and $w_{j_k} \in f^{-1}(w_{j_k}')$.  The result now follows from Lemma 3.4.
\end{proof}

\begin{thm}
Suppose $N=(V,A,r,X)$, $N'=(V',A',r',X)$, and $N''=(V'',A'',r'',X)$ are $X$-networks and $f: N \to N'$ and $g: N' \to N''$ are restricted CSD maps.  Then the composition $g \circ f: N \to N''$ is restricted.
\end{thm}

\begin{proof}
Suppose $v'' \in V''$.  We must show that $(g\circ f)^{-1}(v'') = f^{-1}(g^{-1}(v''))$ is restricted.  Since $g$ is a restricted CSD map,  $g^{-1}(v'')$ is restricted and connected.  Since $f$ is restricted,  $f^{-1}(g^{-1}(v''))$ is also restricted by Lemma 4.4.
\end{proof}

We can now prove the universality property of $ResTr(N)$.  

\begin{thm}
Let $\phi: N \to ResTr(N)$ be the natural projection map.  Then $\phi$ is universal for restricted maps.
\end{thm}

\begin{proof}
Let $g: N \to T$ be a restricted map with kernel $\mathcal{Q}$.  Let $\mathcal{P}$ denote the kernel of $\phi$.  Note that each member $B \in \mathcal{P}$ is restricted and each member $C \in \mathcal{Q}$ is restricted.  

We use Theorem 2.4 to define a CSD map $h: ResTr(N) \to T$ such that $g = h \circ \phi$. We first show that $\mathcal{P}$ is subordinate to $\mathcal{Q}$.  Let $B \in \mathcal{P}$.  We must show that there exists a member $C \in \mathcal{Q}$ such that $B \subseteq C$.  

For any vertex $v$ of $N$ there exists $C(v) \in \mathcal{Q}$ such that $v \in C(v)$ since $\mathcal{Q}$ is a partition.  Since $C(v)$ is restricted,  $R(v) \subseteq C(v)$ by Theorem 3.4.  If $R(v')$ intersects $R(v)$ then $C(v)$ intersects $C(v')$, whence because $\mathcal{Q}$ is a partition it follows $C(v) = C(v')$; hence $R(v) \cup R(v') \subseteq C(v)$.  A simple induction then shows that the member $B \in \mathcal{P}$ that contains $v$ satisfies $B \subseteq C(v)$.   This shows that $\mathcal{P}$ is subordinate to $\mathcal{Q}$.  
\end{proof}

\begin{figure}[!htb]  
\begin{center}

\begin{picture}(350,200) 
\put(130,190){\vector(1,-1){30}}
\put(130,190){\vector(-1,-1){30}}
\put(100,160){\vector(1,-1){30}}
\put(100,160){\vector(-1,-1){30}}
\put(70,130){\vector(-1,0){30}}
\put(70,130){\vector(0,-1){30}}
\put(40,130){\vector(-1,-1){30}}
\put(40,130){\vector(0,-1){30}}
\put(70,100){\vector(-1,-1){30}}
\put(70,100){\vector(1,-1){30}}
\put(100,70){\vector(0,-1){30}}
\put(100,40){\vector(-1,-1){30}}
\put(100,40){\vector(1,-1){30}}
\put(130,130){\vector(1,-1){30}}
\put(130,130){\vector(0,-1){30}}
\put(160,100){\vector(1,-1){30}}
\put(160,100){\vector(0,-1){30}}
\put(130,100){\vector(1,-1){30}}
\put(130,100){\vector(-1,-1){30}}
\put(160,70){\vector(1,-1){30}}
\put(160,70){\vector(0,-1){30}}

\put(170, 160){1}
\put(195, 70){2}
\put(195, 40){3}
\put(160, 30){4}
\put(130, 0){5}
\put(70, 0){6}
\put(40,60){7}
\put(40,90){8}
\put(10,90){9}
\put(140,190){10}
\put(110,160){11}
\put(135,130){12}
\put(160,110){13}
\put(165,70){14}
\put(115,100){15}
\put(110,70){16}
\put(80,100){17}
\put(80,130){18}
\put(110,40){19}
\put(40,140){20}
\put(40,190){$N$}

\put(310,190){\vector(1,-1){30}}
\put(310,190){\vector(-1,-1){30}}
\put(280,160){\vector(0,-1){30}}
\put(280,130){\vector(0,-1){30}}
\put(280,130){\vector(-1,-1){30}}
\put(280,130){\vector(-1,0){30}}
\put(280,130){\vector(1,0){60}}
\put(280,130){\vector(2,-1){60}}
\put(280,130){\vector(1,-1){60}}
\put(280,100){\vector(1,-1){30}}
\put(280,100){\vector(-1,-1){30}}
\put(250,130){\vector(-1,0){30}}
\put(250,130){\vector(-1,-1){30}}

\put(340,165){1}
\put(340,135){2}
\put(340,105){3}
\put(340,75){4}
\put(310,60){5}
\put(250,60){6}
\put(250,90){7}
\put(220,90){8}
\put(220,135){9}
\put(320,190){10}
\put(290,160){11}
\put(285,100){19}
\put(250,135){20}
\put(285,135){$R'(16)$}
\put(250,190){$ResTr(N)$}

\end{picture}

\caption{  An $X$-network $N$ with $X = \{1, 2, 3, 4, 5, 6, 7, 8, 9\}$ and $ResTr(N)$.}  
\end{center}
\end{figure}
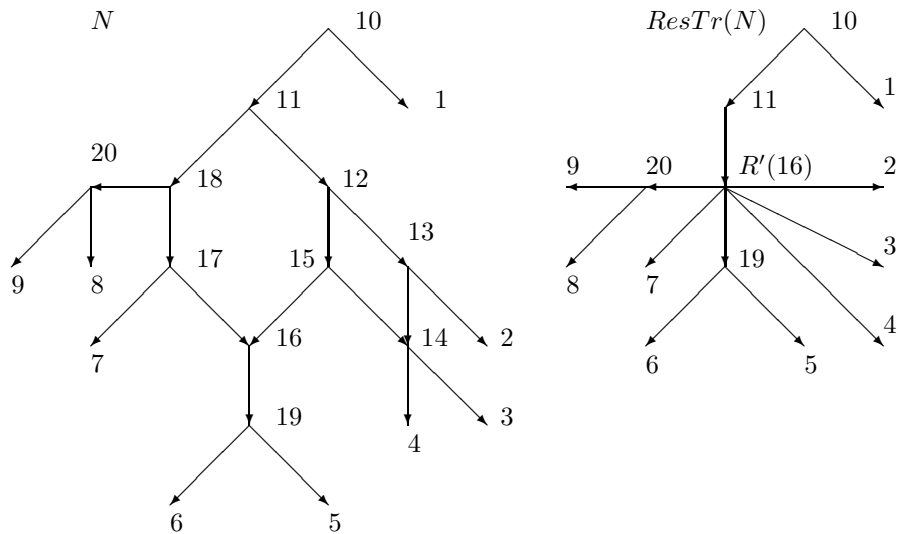

For an example, consider the network $N$ shown in Figure 1.  We demonstrate the construction of $ResTr(N)$, also shown in Figure 1.  Let $v$ be a vertex of $N$.  If $v \notin \{14,16\}$, then $R(v) = \{v\}$ since $v$ has indegree 1.  To compute $R(16)$, initially $R_0 = \{16\}$.  Since $(15,16)$ and $(17,16)$ are arcs and $15  \nleq  17$, we can add 15 to $R_0$ by (2b) yielding $R_1 = \{15,16\}$.  Since $17 \nleq  15$ we can add 17 to $R_1$ by (2a), yielding $R_2= \{15,16,17\}$.  Since $(12,15)$ is an arc and $12 \nleq  17$, we can adjoin 12 by (2a), so $R_3 = \{12,15,16,17\}$.  Since $(18,17)$ is an arc and $18 \nleq  12$, we can adjoin 18 by (2a), so $R_4 = \{12,15,16,17,18\}$.  Now the only arcs $(u,v)$ with $v \in R_4$ and $u \notin R_4$ are $(11,12)$ and $(11,18)$, so we cannot adjoin 11 using (2b).   For all $v \in R_4$, $11\leq v$ so we cannot adjoin 11 by (2a).  Hence the algorithm terminates with $R(16) = R_4 = \{12,15,16,17,18\}$.  

Similarly $R(14) = \{13,14,15\}$.  Since $R(16) \cap R(14) = \{15\}$ is nonempty, $R'(14) = R'(16) = R(14) \cup R(16) = \{12,13,14,15,16,17,18\}$.  For all $v \notin R'(16)$, $R'(v) = R(v) = \{v\}$.  Now $ResTr(N)$ is the quotient digraph.  

As promised, $ResTr(N)$ is a tree.  Note that the resolution of the cluster $\{3,4\}$ in $N$ is lost while that of $\{5,6\}$ is preserved; this is because the hybrid vertex 16 had outdegree 1 while the hybrid vertex 14 had outdegree 2.  The natural projection map $\phi$ takes $\phi(v) = v$ except that for $v \in R'(16)$, $\phi(v) = R'(16)$.   

To illustrate the universality of the map $\phi$ in this example, consider the map $f: N \to T$ where Figure 2 shows $N$ and $T$ in which the vertices of $N$ have been labelled by the vertices of $T$ in order to display the map $f$.  One checks that $f$ is restricted.  For example, $f^{-1}( a)$ is the set of vertices in $N$ labelled $a$ and is restricted.   Then $f$ factors as $f = g \circ \phi$ where $g: ResTr(N) \to T$ satisfies that $g(R'(16)) = a = g(19)$, $g(11) = 10$, $g(20)=b$, and for other vertices $v$ of $ResTr(N)$, $g(v)=v$.

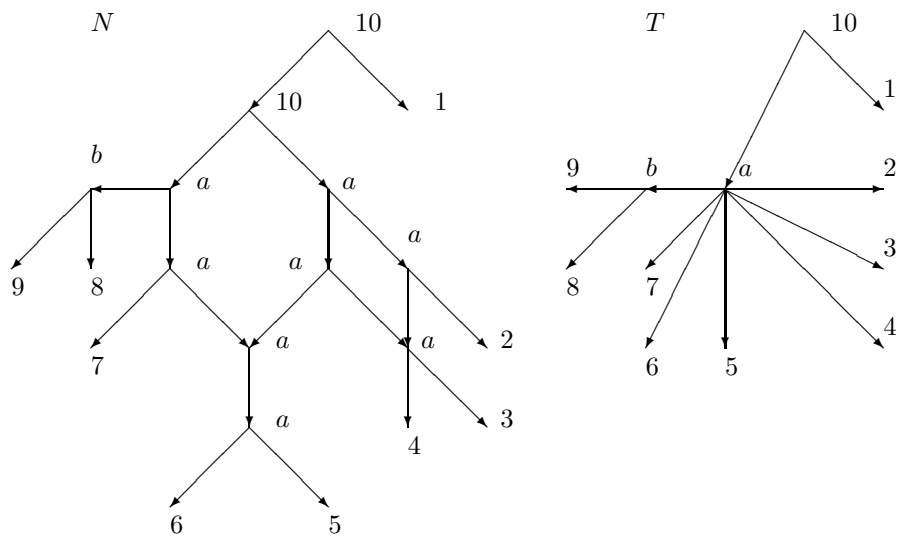
\begin{figure}[!htb]  
\begin{center}

\begin{picture}(350,200) 
\put(130,190){\vector(1,-1){30}}
\put(130,190){\vector(-1,-1){30}}
\put(100,160){\vector(1,-1){30}}
\put(100,160){\vector(-1,-1){30}}
\put(70,130){\vector(-1,0){30}}
\put(70,130){\vector(0,-1){30}}
\put(40,130){\vector(-1,-1){30}}
\put(40,130){\vector(0,-1){30}}
\put(70,100){\vector(-1,-1){30}}
\put(70,100){\vector(1,-1){30}}
\put(100,70){\vector(0,-1){30}}
\put(100,40){\vector(-1,-1){30}}
\put(100,40){\vector(1,-1){30}}
\put(130,130){\vector(1,-1){30}}
\put(130,130){\vector(0,-1){30}}
\put(160,100){\vector(1,-1){30}}
\put(160,100){\vector(0,-1){30}}
\put(130,100){\vector(1,-1){30}}
\put(130,100){\vector(-1,-1){30}}
\put(160,70){\vector(1,-1){30}}
\put(160,70){\vector(0,-1){30}}

\put(170, 160){1}
\put(195, 70){2}
\put(195, 40){3}
\put(160, 30){4}
\put(130, 0){5}
\put(70, 0){6}
\put(40,60){7}
\put(40,90){8}
\put(10,90){9}
\put(140,190){10}
\put(110,160){10}
\put(135,130){$a$}
\put(160,110){$a$}
\put(165,70){$a$}
\put(115,100){$a$}
\put(110,70){$a$}
\put(80,100){$a$}
\put(80,130){$a$}
\put(110,40){$a$}
\put(40,140){$b$}
\put(40,190){$N$}

\put(310,190){\vector(1,-1){30}}
\put(310,190){\vector(-1,-2){30}}
\put(280,130){\vector(0,-1){60}}
\put(280,130){\vector(-1,-1){30}}
\put(280,130){\vector(-1,0){30}}
\put(280,130){\vector(1,0){60}}
\put(280,130){\vector(2,-1){60}}
\put(280,130){\vector(1,-1){60}}
\put(280,130){\vector(-1,0){30}}
\put(280,130){\vector(-1,-2){30}}
\put(250,130){\vector(-1,0){30}}
\put(250,130){\vector(-1,-1){30}}

\put(340,165){1}
\put(340,135){2}
\put(340,105){3}
\put(340,75){4}
\put(280,60){5}
\put(250,60){6}
\put(250,90){7}
\put(220,90){8}
\put(220,135){9}
\put(320,190){10}
\put(250,135){$b$}
\put(285,135){$a$}
\put(250,190){$T$}

\end{picture}

\caption{   A restricted map $f:N \to T$ for $N$ as in Figure 1 is given by the labelling of the vertices of N.  The map factors through $ResTr(N)$.}  
\end{center}
\end{figure}

It is interesting that the vertex 11 in $ResTr(N)$ cannot be removed from $ResTr(N)$ by contracting the arc $(11,R'(16))$ and still retain universality.  In the example of Figure 2, both 11 and 10 in $ResTr(N)$ are mapped to 10 in $T$.  But a simple modification could yield an example in which 11 and 10 in $ResTr(N)$ must go to distinct vertices of the modified $T$.

The network $ResTr(N)$ detects narrow bottlenecks in $N$.  Perhaps it is most appropriate to apply to $ClDis(N)$ (see \cite{wil10}) rather than to $N$ itself, since large regions in $N$ of vertices all with the same cluster can become bottlenecks in $ClDis(N)$.

\end{document}